\documentclass[12pt,a4paper]{article}
\usepackage{epsfig}
\usepackage{amsmath}
\usepackage{graphicx}

\begin{document}

\title{
\hfill {\normalsize MZ-TH/00--54}\\[-3mm]
\hfill {\normalsize Juli 2000}\\[1cm]
Approximate 3-Dimensional Electrical Impedance Imaging}
\author{C. Lehmann\thanks{Supported by Stiftung Rheinland-Pfalz f\"ur
    Innovation} , K.~Schilcher \\ 
{\normalsize {Institut f\"{u}r Physik, Johannes-Gutenberg-Universit\"{a}t,}} 
\\
{\normalsize {Staudinger Weg 7, D-55099 Mainz, Germany}}}
\date{}
\maketitle

\begin{abstract}
\noindent We discuss a new approach to three-dimensional electrical
impedance imaging based on a reduction of the information to be demanded
from a reconstruction algorithm. Images are obtained from a single
measurement by suitably simplifying the geometry of the measuring chamber
and by restricting the nature of the object to be imaged and the information
required from the image. In particular we seek to establish the existence or
non-existence of a single object (or a small number of objects) in a
homogeneous background and the location of the former in the $(x,y)$-plane
defined by the measuring electrodes. Given in addition the conductivity of
the object rough estimates of its\ position along the $z$-axis may be
obtained. The approach may have practical applications.
\end{abstract}

\bigskip

\bigskip

\newtheorem{defin}{Definition} 
\newtheorem{satz}{Satz} 
\newtheorem{lemma}{Lemma} 
\newtheorem{regel}{Regel} 
\newtheorem{behau}{Behauptung} 
\newtheorem{anm}{Anmerkung}

\section{Introduction}

The aim of electrical impedance tomography (EIT)\ is to reconstruct the
conductivity distribution $\sigma (\mathbf{x})$ in the interior of an object 
$\Omega \subset R^{3}$ from electrical measurements on the boundary $%
\partial \Omega $ . For this purpose a number of different current
distributions are applied to the surface of the object via electrodes and
the resulting potentials on the surface are recorded. Applications can be
envisaged both in medicine and industry \cite{ueber}.

Conservation of the current $\mathbf{j}(\mathbf{x})$ and Maxwell's equations
in the quasi-static limit lead to the following differential equation for
the potential $\Phi(\mathbf{x})$: 
\begin{equation}
\mathbf{\nabla}\cdot\lbrack\sigma(\mathbf{x})\mathbf{\nabla}\Phi (\mathbf{x}%
)]\,\,=\,\,0.  \label{dgl}
\end{equation}

In the following we take as the object a rectangular box and investigate
whether statements on the conductivity distribution can be made if the
surface potential can only be measured on one side of the box. Such a model
relates to typical situation in geological and medical imaging.

The general inverse conductivity problem for the box requires current- and
potential-measurements for a large number (in principle infinite) of applied
current configurations on the surface of the box. For the reconstruction of
the conductivity distribution in this and related problems the boundary
conditions must be known precisely and all calculations of potentials be
performed with high accuracy. All these conditions are difficult to be
achieved in practice, which explains the comparative lack of success of the
impedance method in medical applications. In many cases, specifically breast
cancer screening, it is actually not absolutely necessary to have a complete
image of the region. If we restrict the reconstruction to a shadow on a
plane and require only rough information on size and location of the
cancerous region, the reconstruction can be done analytically using a 
\textbf{single} measurement. This problem has also been discussed from
different points of view \cite{ciulli}.

\section{Description of the problem}

We are interested in the conductivity distribution $\sigma(x)$ inside a
rectangular box with sides $a$, $b$, $c$, as pictured in figure
(\ref{kasten}). 
\vspace*{0.5cm} 
\newline

\setlength{\unitlength}{1cm} 
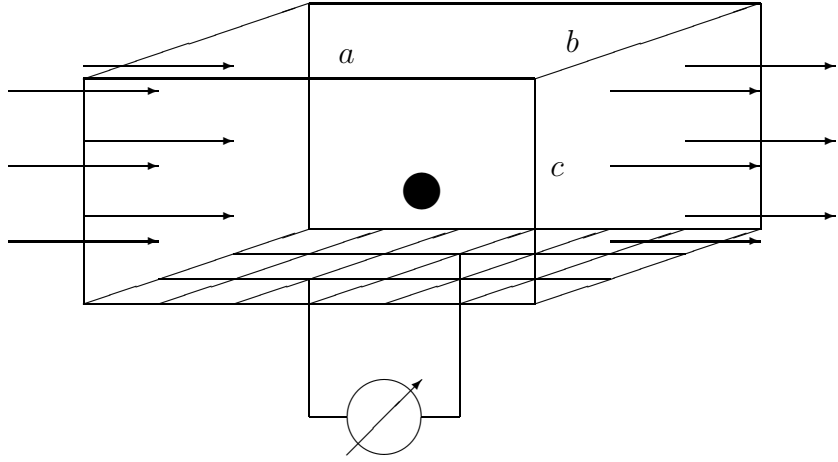
\begin{figure}[th]
\begin{center}
\begin{picture}(12,6)
\put(1,2){\line(1,0){6}}
\put(2,2.3333){\line(1,0){6}}
\put(3,2.6666){\line(1,0){6}}
\multiput(1,2)(1,0){6}{\line(3,1){3}}
\put(1,2){\line(0,1){3}}
\put(4,3){\line(1,0){6}}
\put(4,3){\line(0,1){3}}
\put(7,2){\line(3,1){3}}
\put(7,2){\line(0,1){3}}
\put(10,3){\line(0,1){3}}
\put(1,5){\line(3,1){3}}
\put(1,5){\line(1,0){6}}
\put(4,6){\line(1,0){6}}
\put(7,5){\line(3,1){3}}
\put(0,2.833){\vector(1,0){2}}
\put(0,3.833){\vector(1,0){2}}
\put(0,4.833){\vector(1,0){2}}
\put(1,3.166){\vector(1,0){2}}
\put(1,4.166){\vector(1,0){2}}
\put(1,5.166){\vector(1,0){2}}
\put(8,2.833){\vector(1,0){2}}
\put(8,3.833){\vector(1,0){2}}
\put(8,4.833){\vector(1,0){2}}
\put(9,3.166){\vector(1,0){2}}
\put(9,4.166){\vector(1,0){2}}
\put(9,5.166){\vector(1,0){2}}
\put(4,2.3333){\line(0,-1){1.83333}}
\put(6,2.6666){\line(0,-1){2.16666}}
\put(5.5,0.5){\line(1,0){0.5}}
\put(4.5,0.5){\line(-1,0){0.5}}
\put(5,0.5){\circle{1}}
\put(4.5,0){\vector(1,1){1}}
\put(5.5,3.5){\circle*{0.5}}
\put(4.5,5.3){\makebox(0,0){$a$}}
\put(7.5,5.5){\makebox(0,0){$b$}}
\put(7.3,3.8){\makebox(0,0){$c$}}
\end{picture}
\end{center}
\caption{The geometry of the imaging device}
\label{kasten}
\end{figure}

The region of interest $\Omega$ is therefore of the form 
\begin{equation*}
\Omega=\{(x,y,z)|\,0<x<a,\,0<y<b,\,0<z<c\}.
\end{equation*}

The boundary is made up of six rectangles 
\begin{equation*}
\partial\Omega=\partial\Omega_{x=0}\cup\partial\Omega_{x=a}\cup\partial
\Omega_{y=0}\cup\partial\Omega_{y=b}\cup\partial\Omega_{z=0}\cup\partial
\Omega_{z=c}\;,
\end{equation*}
where, for instance, $\partial\Omega_{x=0}$ means 
\begin{equation*}
\partial\Omega_{x=0}\doteq\{(x,y,z)|x=0,\,0\leq y\leq b,\,0\leq z\leq c\}\;.
\end{equation*}
Similar definitions hold for the other rectangular regions.

The following discussion assumes that a fixed external current enters on one
of the side surfaces and leaves on the opposite surface. The current is
taken to be constant on the two surfaces. i.e. 
\begin{align}
\sigma\frac{\partial\phi}{\partial n}=-I & ;\;\;(x,y,z)\in\partial
\Omega_{x=0}  \label{kr1} \\
\sigma\frac{\partial\phi}{\partial n}=I & ;\;\;(x,y,z)\in\partial
\Omega_{x=a}  \label{kr2} \\
\sigma\frac{\partial\phi}{\partial n}=0 & ;\;\;otherwise,  \label{kr3}
\end{align}
where $\frac{\partial}{\partial n}$ denotes the normal derivative. For
simplicity we set $I=1$, which can always be achieved by a suitable choice
of units for the conductivity and the potential.

We further assume that conditions are such that the resulting potential can
only be measured on the plane $\partial\Omega_{z=0}$ (see fig.(\ref{kasten}%
)).

Given $\sigma(\mathbf{x})$ the resulting potential $\phi(\mathbf{x})$ can be
obtained by solving the differential equation (\ref{dgl}) with the Neumann
boundary condition Eq.(\ref{kr1},\ref{kr2},\ref{kr3}).

The aim is to obtain an image of $\sigma(\mathbf{x})$ from the measurement
potential on the boundary $\partial\Omega_{z=0}$. If the conductivity does
not differ much from a constant distribution $\sigma_{0}$, we can write 
\begin{equation}
\sigma(\mathbf{x})\,\,=\,\,\sigma_{0}\,+\,\delta\sigma(\mathbf{x}).
\end{equation}
Without loss of generality we can set $\sigma_{0}=1$. For $\sigma\equiv
\sigma_{0}$ the solution of the boundary value problem is obviously 
\begin{equation}
\phi_{0}(x,y,z)\,=\,x\,+const.  \label{phic}
\end{equation}

In the following section we try to answer the question to what extend $%
\delta\sigma(\mathbf{x})$ can be reconstructed by measuring the potential
only on the lower surface of the box, i.e. on the boundary surface $%
\partial\Omega_{z=0}$.

\section{Reconstruction}

As the potential distribution is only defined up to a constant, it is
convenient to require that the average of the potential distribution
vanishes on the boundary surface $\partial\Omega$

\begin{equation}
\int_{\partial\Omega}\phi=0  \label{neu1}
\end{equation}

If we assume that the current on the surface $\partial \Omega $ is square
integrable, it is in the space

\begin{equation}
\mathcal{L}_{\diamond}^{2}(\partial\Omega)\,\,:=\,\left\{ f\in\mathcal{L}%
^{2}(\partial\Omega),\,\,\int_{\partial\Omega}\,f\,\,=\,\,0\right\} .
\label{RAUTE}
\end{equation}

Any change $\delta \sigma $ of a homogeneous conductivity distribution $%
\sigma _{0}$ produces a corresponding change $\delta \phi $ in the potential
distribution $\phi _{0}$. Then, for any function $g\in \mathcal{L}_{\diamond
}^{2}(\partial \Omega )$, it can be shown (see appendix), that in linear
approximation 
\begin{equation}
<\delta \phi ,g>_{\mathcal{L}_{\diamond }^{2}(\partial \Omega
)}\,=\,\int_{\partial \Omega }\delta \phi \,g\,\,\,=\,-\int_{\Omega
}\,\nabla \phi _{0}\cdot \nabla \phi _{g}\,\delta \sigma \,\;,
\label{freabl}
\end{equation}
where $\phi _{g}$ represents the solution of the potential problem for
constant conductivity $\sigma _{0}\equiv 1$ and external current
distribution $g$ \cite{hanke}. We have checked in model calculations that
the linearization yields good qualitative images even for objects with large
conductivity. This refers only to the geometrical appearance of the objects
and not to the actual numerical value of the reconstructed conductivity. A
small spherical metallic object of infinite conductivity in a homogeneous
background of conductivity 1 unit, for example, will be imaged as an object
of conductivity 3 units. We will therefore not attempt to determine
numerical values of the conductivity of the hidden objects but only
existence and location of such objects. This restriction of the scope of the
reconstruction will still yield useful results in applications such as
mammography.

For the given experimental set-up we measure a change in potential $\delta
\phi _{exp}$, which we normalize so that it is in $\mathcal{L}_{\diamond
}^{2}(\partial \Omega _{z=0})$, which is defined in analogy to Eq.(\ref
{RAUTE}). We consider a base $\{u_{n}\}$, which is complete and orthonormal
in $\mathcal{L}_{\diamond }^{2}(\partial \Omega _{z=0})$.

It turns out to be useful to introduce in addition a set of functions $%
\tilde{u}_{n}\in\mathcal{L}_{\diamond}^{2}(\partial\Omega)$ (not complete),
which are defined on the full surface $\partial\Omega$ of the box, 
\begin{equation}
\tilde{u}_{n}(\mathbf{x})\,\,:=\,\,\left\{ 
\begin{array}{r@{\quad;\quad}l}
u_{n}(\mathbf{x}) & \mathbf{x}\in\partial\Omega_{z=0} \\ 
0 & else
\end{array}
\right. .  \label{defuns}
\end{equation}
Then, by Eq.(\ref{freabl}), the moments $<\delta\phi_{exp},u_{n} >_{\mathcal{%
L}_{\diamond}^{2}(\partial\Omega_{z=0})}$ satisfy in linear approximation 
\begin{equation}
<\delta\phi_{exp},u_{n}>_{\mathcal{L}_{\diamond}^{2}(\partial\Omega_{z=0}
)}\,\,=\,\,<\delta\phi,\tilde{u}_{n}>_{\mathcal{L}_{\diamond}^{2}
(\partial\Omega)}\,=\,-\int_{\Omega}\,\nabla\phi_{0}\cdot\nabla\phi_{\tilde
{u}_{n}}\,\,\delta\sigma\,.  \label{10}
\end{equation}

We introduce a linear operator $A$ acting on the change in conductivity $%
\delta\sigma$ through 
\begin{equation}
A\,\delta\sigma:=-\sum_{n}\,\left( \int_{\Omega}\,\nabla\phi_{0}\cdot
\nabla\phi_{\tilde{u}_{n}}\,\delta\sigma\,\right) \,u_{n}.  \label{opfisch}
\end{equation}
Using $\delta\phi_{exp}=\sum_{n}<\delta\phi_{exp},u_{n}>_{\mathcal{L}
_{\diamond}^{2} (\partial\Omega_{z=0})}u_{n}$ the relation between $%
\delta\sigma$ and the associated change in potential $\delta\phi_{exp}$ reads

\begin{equation}
A\,\delta \sigma \,=\,\delta \phi _{exp}.  \label{op}
\end{equation}
A natural choice for the base $\{u_{n}\}$ associated to the upper surface is 
\begin{align}
u_{i,j}(x,y)& =C_{i,j}\,\cos {\frac{i\pi x}{a}}\cos {\frac{j\pi y}{b}}%
\,\,\,\,\,\,\,\,i,j=0,1,\cdots ,\,\,\,(i,j)\neq (0,0),  \label{spwa} \\
C_{i,j}& =\left\{ 
\begin{array}{r@{\quad;\quad}l}
2/\sqrt{ab} & i,j\neq 0 \\ 
\sqrt{2/(ab)} & else
\end{array}
\right. ,
\end{align}

where the index $n$ is replaced by two indices $i,j$. The set of functions $%
\tilde{u}_{n}\longrightarrow\tilde{u}_{i,j}$ referring to the whole surface
is then defined in accordance with Eq.(\ref{defuns}). To make use of Eq.(\ref
{freabl}) to calculate $\delta\sigma$ we need the potential $\phi_{\tilde{u}%
_{i,j}}$ resulting from an external current distribution $\tilde{u}_{i,j}\in%
\mathcal{L}_{\diamond}^{2}(\partial\Omega)$ and conductivity $%
\sigma_{0}\equiv1$. It is a simple exercise to show that 
\begin{equation}
\phi_{\tilde{u}_{i,j}}(x,y,z)\,=\,\frac{C_{i,j}}{\delta_{i,j}(1-e^{-2\delta
_{i,j}c})}\cos{\frac{i\pi x}{a}}\cos{\frac{j\pi y}{b}}\left\{ e^{-\delta
_{i,j}z}+e^{\delta_{i,j}z-2\delta_{i,j}c}\right\} ,  \label{pof}
\end{equation}

with the abbreviation 
\begin{equation*}
\delta_{i,j}=\pi\sqrt{(i/a)^{2}+(j/b)^{2}}.
\end{equation*}

If we define 
\begin{align}
\sigma_{i,j} & =||\nabla\phi_{0}\cdot\nabla\phi_{\tilde{u}_{i,j} }||  \notag
\\
& =\frac{i\,\pi}{a\delta_{i,j}(1-e^{-2\delta_{i,j}c})}\left( \frac
{1}{2\delta_{i,j}}\left( 1-e^{-4\delta_{i,j}c}\right) +2c\,e^{-2\delta_{i,j}
c}\right) ^{1/2},  \label{sigmaij} \\
v_{i,j} & =\frac{-\nabla\phi_{0}\cdot\nabla\phi_{\tilde{u}_{i,j}}}{\|
\nabla\phi_{0}\cdot\nabla\phi_{\tilde{u}_{i,j}}\|_{\mathcal{L}^{2} (\Omega)} 
}  \notag \\
& = \,C_{i,j}\sin{\frac{i\pi x}{a}}\cos{\frac{j\pi y}{b}}\left\{
e^{-\delta_{i,j}z}+e^{\delta_{i,j}z-2\delta_{i,j}c}\right\}  \notag \\
& \times\left( \frac{1}{2\delta_{i,j}}\left( 1-e^{-4\delta_{i,j}c}\right)
+2c\,e^{-2\delta_{i,j}c} \right) ^{-1/2},  \label{vij}
\end{align}
then Eq.(\ref{opfisch}) can be written in the form

\begin{equation}
A\,\delta \sigma \,\,=\,\,\sum_{i=1,\,j=0}^{\infty }\,\sigma
_{i,j}\,\,<\delta \,\sigma ,v_{ij}>_{\mathcal{L}^{2}(\Omega )}\,u_{ij}.
\label{ads}
\end{equation}
This is our main result. It is obvious from Eq.(\ref{ads}) that the set $%
\{v_{i,j}\}$ is a complete orthogonal system in $N(A)^{\perp }$. We have
thus explicitly constructed the singular system $\{v_{i,j},u_{i,j};\sigma
_{i,j}\}$ of the operator $A$ and its generalized inverse can be written
down explicitly. The generalized or least square solution of Eq.(\ref{ads})
is then simply given by 
\begin{equation}
\delta \sigma \,=\,\sum_{i=1,\,j=0}^{\infty }\,\sigma _{i,j}^{-1}\,<\delta
\phi _{exp},u_{i,j}>\,v_{i,j}.  \label{va}
\end{equation}
This generalized solution is still not continuous in the data and must be
regularized in a suitable manner. We employed for convenience mainly the
method of truncating the singular values or truncating the indices $i,j$ in
Eq.(\ref{va}). The latter procedure turns out to produce better images. The
cut off values of the indices is determined by a version of the discrepancy
principle, i.e. by requiring that the resolution implied by the Fourier
series Eq.\ref{spwa} should not exceed the distance between the electrodes
which measure the potential on the surface $\partial \Omega _{z=0}$. We
assume the latter constitutes the main source of the experimental error.

It should be pointed out that we only reconstruct a three dimensional
picture which is a projection on the set of functions $\{v_{i,j}\}$. It is
sufficient to view the image at $z=0$, because the images for $z\ne0$ follow
uniquely from the $z=0$ one and contain no additional information.

We effectively see a two-dimensional image, which represents a kind of
shadow of the object.  For many purposes (such as in cancer screening), when
one is only interested in the presence or absence of an object, this is
sufficient information. As discussed above, the actual value of the object's
conductivity cannot be reconstructed quantitatively.

\begin{figure}[th]
\begin{center}
\begin{picture}(14,7)
\put(0,0){\epsfig{file=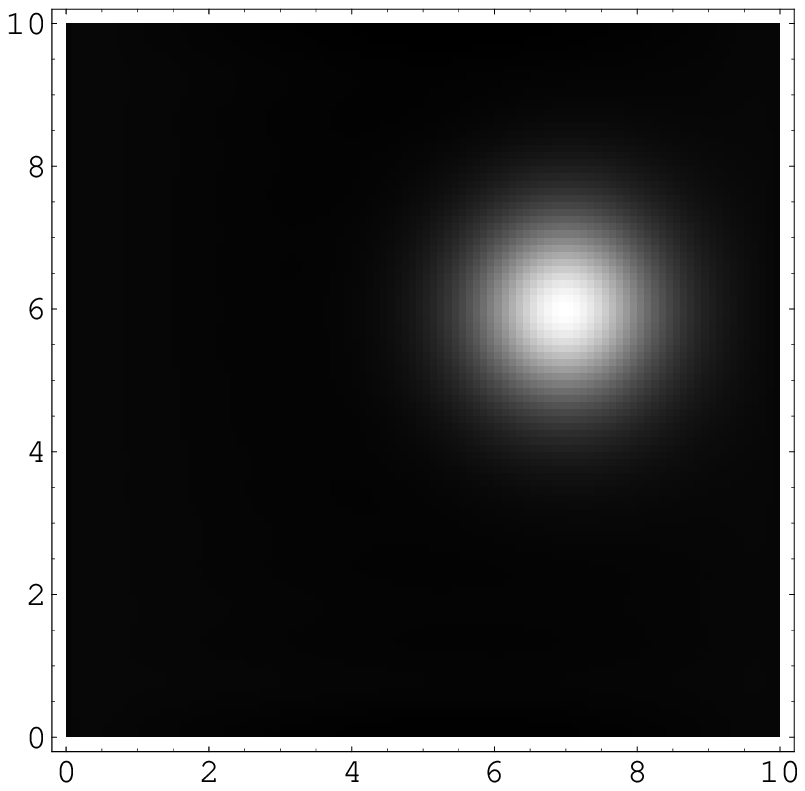,width=6.5cm}}
\put(7,0){\epsfig{file=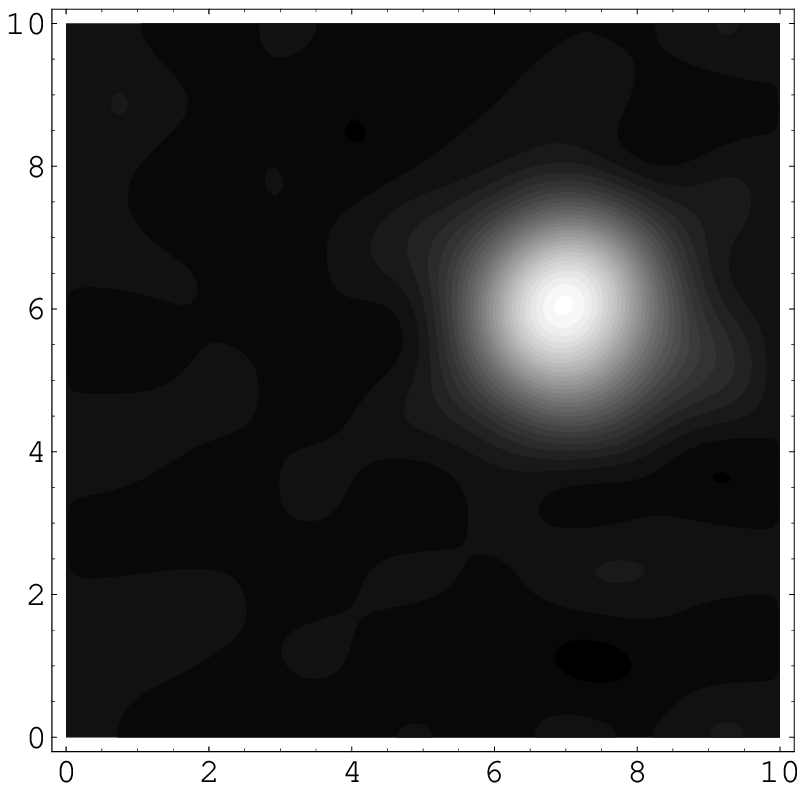,width=6.5cm}}
\put(3.5,6.9){\makebox(0,0){a}}
\put(10,6.9){\makebox(0,0){b}}
\end{picture}
\end{center}
\caption{Images of an spherical object obtained from exact and error
affected data}
\label{bild1}
\end{figure}

\begin{figure}[th]
\begin{center}
\begin{picture}(14,7)
\put(0,0){\epsfig{file=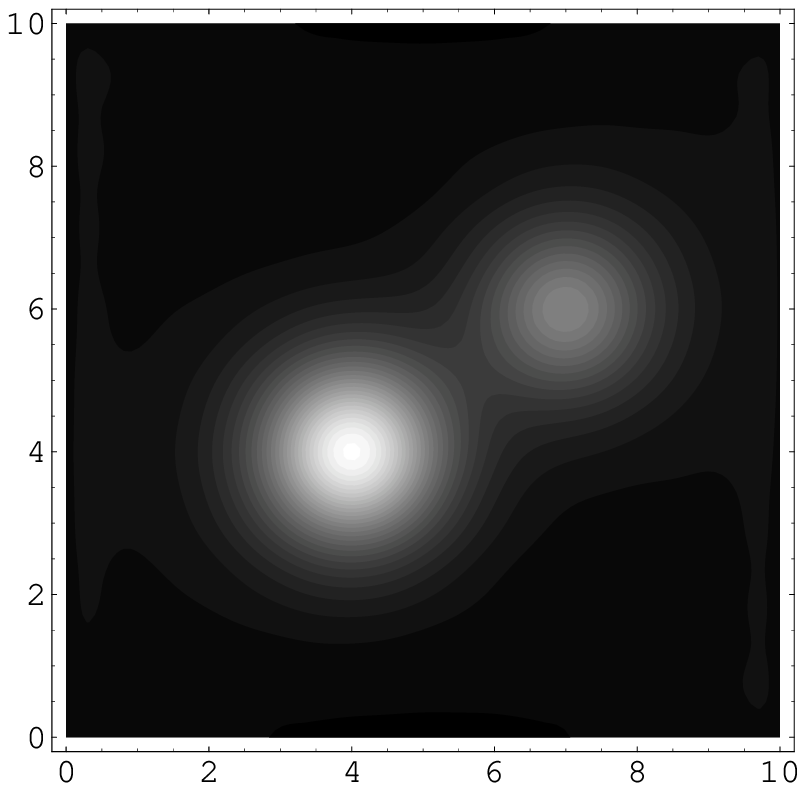,width=6.5cm}}
\put(7,0){\epsfig{file=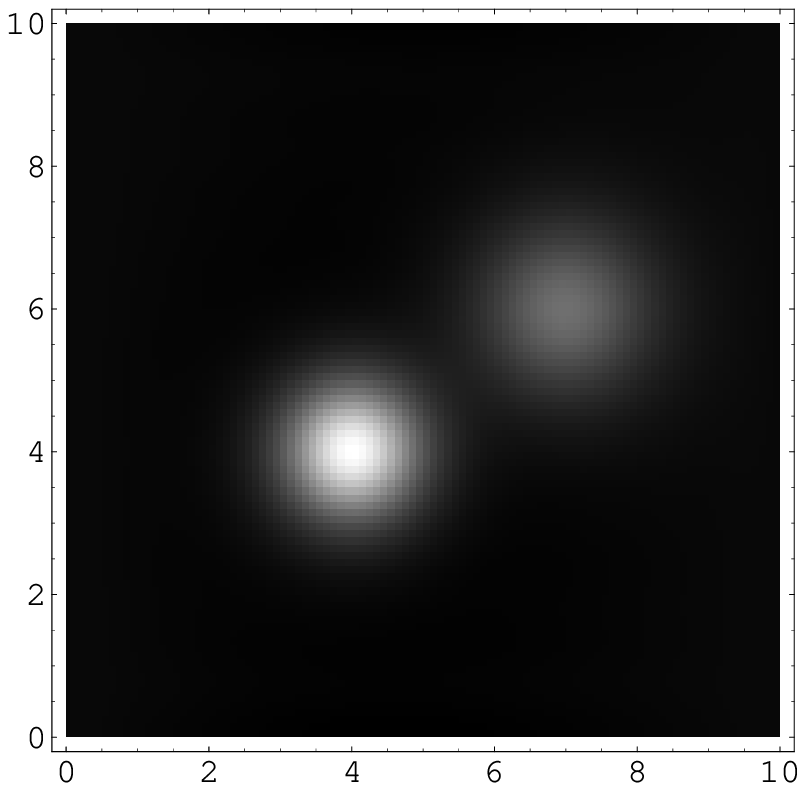,width=6.5cm}}
\put(3.5,6.9){\makebox(0,0){a}}
\put(10,6.9){\makebox(0,0){b}}
\end{picture}
\end{center}
\caption{Images of two spherical objects}
\label{bild2}
\end{figure}

In Fig.(\ref{bild1}) and (\ref{bild2}) we present images obtained from
synthetic data which were calculated for $10\times 10$ grid points on the
plane $\partial \Omega _{z=0}$ . We also show how the image deteriorates
when errors are assigned to the data.

\noindent Figure (\ref{bild1}) shows images of a spherical metallic object
of diameter $d=1$ at a distance $z=2$ (all in units of the grid spacing)
from the surface of measurement, (a) with exact data, and (b) with data $%
\delta \phi _{exp}$ corrupted with a 20\% random uniform multiplicative
error. It is amazing that even with errors of such a magnitude a reasonable
image is produced.

\noindent Figure (\ref{bild2}) shows images of two spherical objects
obtained from exact data. Case (a) shows the image for two spheres of
diameter $1.5$ and $1$ respectively both at a distance $z=2$ from the
surface and case (b) shows the image for two spheres of equal diameter at
distance $z=1.5$ and $z=2$ respectively.

Qualitatively the image gets larger and flatter as the object is moved away
from the measuring plate. In addition the image gets brighter (but not
larger!) as the object gets larger. The same effect is observed when the
conductivity is increased. It is not possible to distinguish volume- from
conductivity effects. This is true as long as the objects are small or not
to close to the surface, as they effectively behave as dipoles (see section
4). Given additional information, e.g. that the object's conductivity is
constant and of a given magnitude, it may be possible to quantify this
observation and obtain a full three dimensional image of the object. This is
exemplified in the next section.

\section{Spherical Object}

In the following we consider a single spherical object $K$ of conductivity $%
\kappa $ and radius $a$ immersed in the box $\Omega $ filled with a liquid
of conductivity $1.$

Let $n_{K}$ be the normal to the surface of $K$ and $n_{\Omega}$ the normal
on $\partial\Omega$. The boundary value problem for a current distribution $%
f\in\mathcal{L}_{\diamond}^{2}(\partial\Omega)$ can be defined as follows, 
\begin{align}
\triangle\phi(x) & =0\,\,\,\,,x\in\Omega\setminus\partial\,K,
\label{difrac1} \\
\frac{\partial\phi}{\partial\,n_{\Omega}}(x) & =f(x)\,\,\,\,,x\in
\partial\Omega,  \label{difrac2} \\
\lim_{h\rightarrow0^{+}}\,(\phi(x+h\,n_{K})-\phi(x-h\,n_{K})) &
=0\,\,\,\,,x\in\partial K,  \label{difrac3} \\
\lim_{h\rightarrow0^{+}}\,\left( \frac{\partial\phi(x+h\,n_{K})}{\partial
n_{K}}-\kappa\frac{\partial\phi(x-h\,n_{K})}{\partial n_{K}}\right) &
=0\,\,\,\,,x\in\partial K,  \label{difrac4} \\
\int_{\partial\Omega}\,\phi\,ds & =0.  \label{difrac5}
\end{align}
Equation (\ref{difrac3}) guarantees the continuity of the potential while (%
\ref{difrac4}) describes current conservation. These two equations determine
the boundary conditions on the surface of the sphere. The other three
equations represent the well-known boundary value problem of the Laplace
equation. The Neumann boundary condition (\ref{difrac2}) is given in (\ref
{kr1},\ref{kr2},\ref{kr3}). For $\kappa=1$ one obtains $\phi_{0}$, the
solution of (\ref{phic}).

For the case of a small sphere of constant conductivity and not too close to
the surface $\Omega ,$ the change in potential $\delta \phi $ is given by
the dipole term, 
\begin{equation}
\delta \phi (\vec{x})=-\frac{\alpha a^{3}\,\nabla \phi _{0}\cdot (\vec{x}-%
\vec{x}_{0})}{|\vec{x}-\vec{x}_{0}|^{3}}\;;  \label{20}
\end{equation}
where $\vec{x}_{0}$ is the coordinate of the centre of the sphere, and

\begin{equation}
\alpha \,\,=\,\,\frac{\kappa -1}{\kappa +2}  \label{21}
\end{equation}
This result can be derived by noting that our boundary value problem is
equivalent to that of a dielectric sphere in a uniform electric field.

The variation of $\alpha $ with $\kappa $ shows quite clearly the limited
sensitivity of EIT to changes of conductivity. In practical reconstructions $%
\kappa =10$ can hardly be distinguished from $\kappa =\infty .$

The potential $\phi =\phi _{0}+\delta \phi $ still does not satisfy the
Neumann boundary condition (\ref{difrac2}) on $\partial \Omega $. This
problem can, in principle, be solved by an infinite number of image dipoles.
We checked that the series converges rapidly. For the case $a,b\rightarrow
\infty $ , i.e. the case of two infinite plates,the sum takes on the simple
form:

\begin{align}
\delta \phi (\vec{x})& =-\alpha r^{3}\sum_{n=0}^{\infty }\left\{ \frac{%
\nabla \phi _{0}\cdot (x-x_{0},y-y_{0},z-(2nc-z_{0}))^{T}}{%
|(x-x_{0})^{2}+(y-y_{0})^{2}+(z-(2nc-z_{0}))^{2}|^{3}}+\right.  \notag \\
& +\left. \frac{\nabla \phi _{0}\cdot (x-x_{0},y-y_{0},z-(-2nc+z_{0}))^{T}}{%
|(x-x_{0})^{2}+(y-y_{0})^{2}+(z-(-2nc+z_{0}))^{2}|^{3}}\right\}  \label{22}
\end{align}

The knowledge of $\delta \phi $ allows to calculate the generalized inverse
according to Eq.(\ref{va}). For rough estimates and when the sphere is not
to close to the surface, the image dipoles can be neglected. As the
conductivity of the sphere is assumed to be known, Eq.(\ref{20}) can be used
to obtain an estimate of the position and the volume of the sphere. As a
test we measure the synthetic data on the surface $\partial \Omega _{z=0}$
by $10\times 10$ electrodes. Given the rough knowledge of the coordinates $%
(x_{0},y_{0})$ of the centre of the sphere, we fit $\delta \phi $ of Eq.(\ref
{20}) plus the background potential $\phi _{0}$ to data taken on neighboring
electrodes. As a typical example, we find for data afflicted with 10\%
multiplicatice uniform error, the following results: $z_{0}=2.32$ (instead
of $2.0$) and $r=0.56$ (instead of $0.5$).

In realistic applications the object to be detected will in general not be
spherical. Nonetheless one may obtain rough information on size and depth of
the location of the object by assuming a spherical shape and applying the
analysis above.

\section{Conclusion}

We have presented in this note an electric impedance imaging system based on
a specific simple geometry of the device which guarantees a uniform current
distribution in the case of constant conductivity. If we further impose the
condition that only a single object (or possible a small number of objects)
is to be detected, then we show that an image can be obtained in a single
measurement of the surface potential. To test of the effectiveness of the
method, we create synthetic data which can be afflicted with errors. The
image obtained by inverse problem techniques represents a projection or
shadow on the surface where the potential is measured. This image is
amazingly stable against data errors. We also indicate how rough estimates
on the size and the depth of the object may be obtained. The actual
construction of such an imaging system is planned.

\appendix

\section{Appendix}

We will give the sketch of a proof of Eq.(\ref{freabl}). Let a potential,
denoted by $u(\mathbf{x})$, satisfy the EIT differential equation 
\begin{equation}
\,\mathbf{\nabla (}\sigma \mathbf{\nabla }u(\mathbf{x))=0\;}  \label{a0}
\end{equation}
for a given conductivity $\sigma (\mathbf{x)}$ and surface current $f(%
\mathbf{x})$ 
\begin{equation}
f=\sigma \left. \frac{\partial u}{\partial n} \right|_{\partial \Omega }\;\;.
\label{a1}
\end{equation}
Let $v(\mathbf{x})$ be an arbitrary solution of the EIT differential
equation Eq.\ref{a0}. Then we can define a functional 
\begin{equation}
b_{f}[v]:=\int_{\partial \Omega }f\cdot v  \label{a2}
\end{equation}
and a bilinear form. 
\begin{equation}
a_{\sigma }[u,v]:=\int_{\Omega }\sigma \,\mathbf{\nabla }u\cdot \mathbf{%
\nabla v\;\;.}  \label{a3}
\end{equation}
The EIT boundary value problem with Neumann boundary conditions is \ known $
\cite{mikhlin}$to be equivalent to the condition 
\begin{equation}
b_{f}[v]=a_{\sigma }[u,v]\;\;\forall v  \label{a4}
\end{equation}
We now change $\sigma \rightarrow \sigma +\delta \sigma $, keeping the same
Neumann boundary condition \ref{a1}. Then the potential will change as $%
u\rightarrow u+\delta u$ and the condition Eq.\ref{a4} will read 
\begin{equation}
b_{f}[v]=a_{\sigma +\delta \sigma }[u+\delta u,v]\;\;\forall v  \label{a5}
\end{equation}
Or 
\begin{equation}
b_{f}[v]=a_{\sigma }[u,v]+a_{\sigma }[\delta u,v]+a_{\delta \sigma
}[u,v]+a_{\delta \sigma }[\delta u,v]  \label{a6}
\end{equation}
Neglecting \ the last term,using Eq.\ref{a4} and the symmetry of the
bi-linear form, we obtain the relation 
\begin{equation}
a_{\delta \sigma }[u,v]=-a_{\sigma }[v,\delta u]\;\;\forall v  \label{a7}
\end{equation}
As $v(\mathbf{x})$ is the solution of Eq.\ref{a0} for some boundary current $%
g(\mathbf{x})$, it must satisfy 
\begin{equation}
b_{g}[w]=a_{\sigma }[v,w]  \label{a8}
\end{equation}
for all $w(\mathbf{x})$, in particular for $w(\mathbf{x})=\delta u(\mathbf{x}%
)$. We finally obtain therefore 
\begin{equation}
\int_{\partial \Omega }g\cdot \delta u=-a_{\delta \sigma }[u,v]\;,
\label{a9}
\end{equation}
which is just Eq.(9).

\end{document}